\documentclass[reprint,amsmath,amssymb,aps,pra,floatfix]{revtex4-2}

\usepackage{amsfonts}
\usepackage{amsmath}
\usepackage{amsthm}
\usepackage{mathrsfs}
\usepackage{amscd}
\usepackage{amssymb}
\usepackage{subfigure}
\usepackage{amsxtra}
\usepackage{dcolumn}
\usepackage{bm}           
\usepackage{bbm}
\usepackage{graphicx}
\usepackage{epstopdf}

\usepackage[colorlinks]{hyperref}

\newcommand{\la}{\langle}
\newcommand{\ra}{\rangle}

\newcommand{\da}{\dagger}

\newcommand{\Op}[1]{\hat{#1}}

\newcommand{\oH}{\Op{H}}

\newcommand{\oU}{\Op{U}}

\newcommand{\oV}{\Op{V}}

\renewcommand{\vr}{\boldsymbol{r}}

\newcommand{\tr}{\ensuremath{{\rm tr}}}

\newcommand{\diff}{\mathrm{d}}

\DeclareMathOperator{\var}{Var}

\begin{document}

\preprint{APS/123-QED}

\title{Exact Entanglement Dynamics of Two Spins in  Finite Baths}
\author{Mei Yu}
\affiliation{Naturwissenschaftlich-Technische Fakult{\"a}t, Universit{\"a}t Siegen, Siegen 57068, Germany}
\author{Otfried Gühne}
\affiliation{Naturwissenschaftlich-Technische Fakult{\"a}t, Universit{\"a}t Siegen, Siegen 57068, Germany}
\author{Stefan Nimmrichter}
\affiliation{Naturwissenschaftlich-Technische Fakult{\"a}t, Universit{\"a}t Siegen, Siegen 57068, Germany}

\date{\today}
             
\begin{abstract}
We consider the buildup and decay of two-spin entanglement through phase interactions in a finite environment of surrounding spins, as realized in quantum computing platforms based on arrays of atoms, molecules, or nitrogen vacancy centers. The non-Markovian dephasing caused by the spin environment through Ising-type phase interactions can be solved exactly and compared to an effective Markovian treatment based on collision models.
In a first case study on a dynamic lattice of randomly hopping spins, we find that non-Markovianity boosts the dephasing rate caused by nearest neighbour interactions with the surroundings, degrading the maximum achievable entanglement. However, we also demonstrate that additional three-body interactions can mitigate this degradation, and that randomly timed reset operations performed on the two-spin system can help sustain a finite average amount of steady-state entanglement.
In a second case study based on a model nuclear magnetic resonance system, we elucidate the role of bath correlations at finite temperature on non-Markovian dephasing. They speed up the dephasing at low temperatures while slowing it down at high temperatures, compared to an uncorrelated bath, which is related to the number of thermally accessible spin configurations with and without interactions.
\end{abstract}
\maketitle

\section{Introduction}
Quantum computers hold the promise of being one of the next major technological developments in the field of information technology \cite{Nielsen2002, Ladd2010}. Most of them operate on physical platforms realizing discrete arrays of physical qubits that can be addressed individually and intercoupled with others in their vicinity. 
Quantum entanglement not only provides these platforms with the ability to potentially solve 
hard computational problems and simulate quantum systems more efficiently than classical computers \cite{Arute2019,Bennett2000}, but also facilitates the redundant encoding of logical quantum states into multiple physical qubits, known as quantum error correcting codes \cite{shor1995,Knill1997,Knill2000,Yao2012,Brooks2013,Terhal2015}. 
Both use cases are based on methods to reliably generate and preserve entanglement between neighbouring physical qubits in the presence of unavoidable noise and decoherence from their surroundings. 

Entanglement can deteriorate due to technical noise from applied control fields and also due to dephasing caused by unwanted residual interactions with other quantum systems nearby. Understanding decoherence and protecting the entanglement of quantum systems is a central challenge in quantum science and technology. Existing strategies to suppress decoherence and stabilize most of the generated entanglement include dynamical decoupling by time-modulated control fields \cite{Duan1997,Viola1998, Vitali1999,Viola1999, Agarwal2001} and the use of decoherence-free subspaces \cite{Lidar1998,Lidar2014}. 
A much simpler scheme is based on sequences of reset operations to counteract environmental dephasing and thereby uphold a steady state containing a usable fraction of the ideally generated entanglement \cite{Hartmann2006}. A reset operation replaces the reduced state of a system coupled to its surroundings with a freshly prepared fiducial state. Stochastic sequences of such resets were also studied in the context of classical diffusion processes \cite{Evans_2020,Evans2011,Evans_2011,Montero2013}.

In this paper, we introduce a generic spin model to study the buildup and decay of entanglement through controlled phase rotations between spins subjected to dephasing of many surrounding spins arranged, e.g., on a lattice. This may represent scenarios of quantum state processing on an array of Rydberg atoms \cite{Christian2017,Bernien2017,Browaeys2016,Barredo2018}, selected constituent nuclear spins of a molecule \cite{Niknam2021}, an array of trapped polar molecules \cite{DeMille2002, Yelin2006, Wei2016, Loic2019, Hughes2020}, a hybrid array of molecules and atoms \cite{Wang2022, Zhang2022}, or also a system of nitrogen vacancy (NV) centers \cite{Jelezko22004,Robledo2011,Aboebeih2018,Degen2021,Gulka2021,Maile2022}. 
By allowing the spins to also hop through the lattice as in Refs.~\cite{Calsamiglia2005,Hartmann2005}, we can directly compare the effects of Markovian and non-Markovian dephasing on the achievable entanglement. Our general finding is that non-Markovianity accelerates the entanglement decay compared to the Markovian case. However, we also show that a substantial degree of steady-state entanglement can be preserved by making use of random reset operations on the system spins. 
We also introduce a three-spin phase interaction between system and environment, which can partly alleviate the dephasing effect. 

In addition, we investigate the role of initial environment correlations and temperature by considering a finite thermal bath of interacting spins in a case study of nuclear magnetic resonance (NMR) processing of two central spins of a single molecule. The bath correlations turn out to be detrimental as they result in low-energy excitations that enhance the dephasing effect at low temperatures, while they reduce it at higher temperatures relative to an uncorrelated spin bath.

The paper is organized as follows: Section \ref{sec:theory} introduces our model blueprint for interacting spins subject to dephasing in a static or evolving spin environment. In Section \ref{sec:lattice}, a case study on entanglement generation by a controlled phase gate, we elucidate the difference between Markovian and non-Markovian dephasing in a spin lattice with nearest-neighbour phase interactions. We also assess the amount of steady-state entanglement one can achieve with help of random reset operations. Three-body interactions are also discussed in this section as a means to alleviate the dephasing process. Section \ref{sec:molecule} proceeds with a scenario of a static correlated spin environment with long-range interactions, representing a molecule. Finally, we conclude in Section \ref{sec:conclusion}.

\section{Theoretical model}
\label{sec:theory}

\begin{figure}[t]
    \centering
     \includegraphics[width=\linewidth]{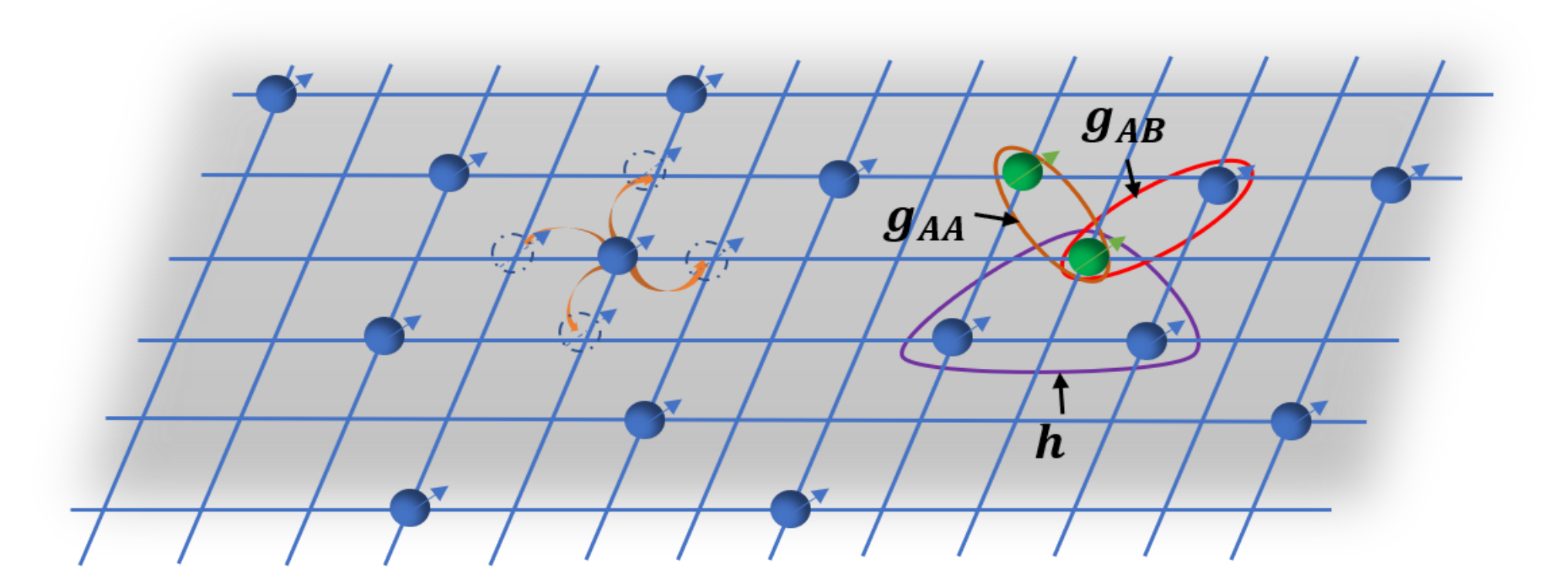}
    \caption{Dynamical spin network model for non-Markovian dephasing. We consider a system of $N_A$ elementary quantum spins (green, $N_A=2$) surrounded by $N_B$ environmental spins (blue), here occupying sites of a regular periodic lattice. The configuration can be static or dynamic, depending on whether the spins can move on the lattice. The spins exchange phase information via short- or long-range Ising-type interactions. We distinguish between intra-system and system-environment couplings illustrated by the links of strength $g_{AA}$ and $g_{AB}$, respectively. We also consider three-body phase interactions of strength $h$ between neighbours, as marked by the triplet.  
    }
    \label{fig:sketch}
\end{figure}

As a scheme for spin dephasing in a finite-size environment, we consider a generic spin configuration such as the the periodic lattice model introduced in Refs.~\cite{Calsamiglia2005,Hartmann2005} and depicted in Fig.~\ref{fig:sketch}, in which $N=N_A+N_B$ spins occupy some or all available sites. We distinguish a small number of $N_A$ accessible `system' spins (here $N_A=2$) and $N_B \gg 1$ `environment' spins, which either remain static or are allowed to hop between sites. The motion is described through time-dependent (discrete) position vectors $\vr_n (t)$, with $n=1,\ldots, N$. 
In our study, we will consider two exemplary scenarios: a two-dimensional partially filled lattice like the depicted one, with hopping and nearest-neighbour interactions, and a practical rigid molecular structure, with fixed positions and distance-dependent long-range interactions.

For the local spin energies and interactions, we consider an Ising-like Hamiltonian of the general form
\begin{equation}
\begin{split}
    \oH_{\alpha} (t) &= \sum_n \omega (\bm r_n(t)) |1_n\ra\la 1_n| + \sum_{k<n} g(\bm r_k(t), \bm r_n(t)) \\
    & \left( |1_k \ra \la 1_k |-\frac{\alpha}{2}\openone_k \right) \otimes \left( |1_n \ra \la 1_n | - \frac{\alpha}{2}\openone_n \right)     \label{eq:H} \\
    & + \sum_{\ell<k<n} h (\bm r_\ell(t),\bm r_k(t),\bm r_n(t)) \left( |1_\ell \ra \la 1_\ell |- \frac{\alpha}{2}\openone_\ell \right)\\ 
    &\otimes \left( |1_k \ra \la 1_k | - \frac{\alpha}{2}\openone_k \right) \otimes \left( |1_n \ra \la 1_n | -\frac{\alpha}{2}\openone_n \right), 
\end{split}
\end{equation} 
with $|0_n\ra,|1_n\ra$ and $\openone_n$ the computational $Z$-basis states of the $n$-th spin and the identity, respectively. 
We are mainly concerned with homogeneous lattices in which the local spin energies $\omega (\bm r)$ are all the same and can thus be omitted in a co-rotating frame. 
The parameter $\alpha$ determines how the spin interactions modulate the overall energy level spectrum, and we shall distinguish two relevant cases: $\alpha=0$ in Rydberg systems where each spin represents two atomic levels with dipole-dipole Rydberg blockade interactions, and $\alpha=1$ in physical spin models with Ising interactions between spin-$Z$ components.
Conveniently regrouping the terms and shifting the energy zero point, we can write the Hamiltonian as
\begin{equation}
\begin{split}
    \oH_{\alpha} (t) &= \sum_n \tilde{\omega}_n (t) |1_n\ra\la 1_n|  + \sum_{k<n} \tilde{g}_{kn} (t) |1_k 1_n \ra \la 1_k 1_n | \\
    & + \sum_{\ell<k<n} h_{\ell k n} (t) |1_\ell 1_k 1_n \ra \la 1_\ell 1_k 1_n |, \label{eq:Halt}
\end{split}
\end{equation}
with $h_{\ell k n} (t) = h (\bm r_\ell(t),\bm r_k(t),\bm r_n(t))$ and the renormalized one- and two-spin terms
\begin{equation}
    \begin{split}
       \tilde{g}_{kn} (t) &= g(\bm r_k(t), \bm r_n(t)) - \frac{\alpha}{2} \sum_{\ell \neq k,n} h (\bm r_\ell(t),\bm r_k(t),\bm r_n(t)), \\
    \tilde{\omega}_n (t) &= \omega (\bm r_n(t)) - \frac{\alpha}{2} \sum_{k \neq n}  \Bigg[ g(\bm r_k(t), \bm r_n(t)) \\
    &- \frac{\alpha}{2} \sum_{\ell \neq k,n} h (\bm r_\ell(t),\bm r_k(t),\bm r_n(t))  \Bigg].  
    \end{split}
    \label{eq:rescaled_w_g}
\end{equation}

Notice that the $h_{\ell k n}$ and the $\tilde{g}_{kn}$ to be symmetric under permutation of the indices.
Since all terms in the Hamiltonian commute with each other, we can express the corresponding unitary time evolution as
\begin{equation}
    \begin{split}
    \oU(t) = &\prod_n e^{-i \int_0^t \diff t' \, \tilde{\omega}_n (t') |1_n \ra\la 1_n| } \\
    & \prod_{k < n} e^{-i \int_0^t \diff t' \, \tilde{g}_{kn} (t') |1_k 1_n \ra\la 1_k 1_n| } \\
    &\prod_{\ell < k} e^{-i \int_0^t \diff t' \, h_{\ell k n} (t') |1_\ell 1_k 1_n \ra\la 1_\ell 1_k 1_n| },
    \label{eq:U}
    \end{split}
\end{equation}
where the accumulated one-, two-, and three-body phases are determined by each qubit's hopping trajectory from site to site. In the case of static positions, all accumulated phases are linear in $t$.
The first term corresponds to local phase rotations, which do not affect the entanglement. We can absorb them by switching to the rotating frame,
\begin{equation}
    \begin{split}
    \oU_t = & \prod_{k < n} e^{-i \int_0^t \diff t' \, \tilde{g}_{kn} (t') |1_k 1_n \ra\la 1_k 1_n| } \\
    &\prod_{\ell < k < n} e^{-i \int_0^t \diff t' \, h_{\ell k n} (t') |1_\ell 1_k 1_n \ra\la 1_\ell 1_k 1_n| } =: \oU_t^{(2)} \oU_t^{(3)},
    \end{split}
\end{equation}
where $\oU_t^{(2)}$ and $\oU_t^{(3)}$ subsume the two- and three-body terms, respectively.
As a convenient notation, we introduce binary arrays
$\underline{a} = \left(a_1,\ldots, a_{N_A}\right) \in \{0,1\}^{\times N_A}$, $\underline{b} \in \{0,1\}^{\times N_B}$, and the concatenated array $\underline{ab}$ to denote the spin configuration of the energy eigenstates of system and environment,
\begin{equation}
    |\underline{ab} \ra = |\underline{a}\ra | \underline{b}\ra = |a_1, \ldots ,a_{N_A}, b_1, \ldots, b_{N_B} \ra \in \mathbb{C}^{2(N_A+N_B)},
\end{equation}
where the inner products $\underline{a} \cdot \underline{a}$ and $\underline{b} \cdot \underline{b}$ give the numbers of $A$- and $B$-spins in the state $|1\ra$, respectively.

Concerning the pairwise phase interactions, we distinguish between a fixed intra-system coupling strength that realizes a phase gate for entanglement generation, and (smaller) position dependent system-environment coupling strengths constituting the dephasing. 
The three-spin interactions between the system ($N_A=2$) and the environment can either amplify or mitigate the dephasing effect, as we will see in Sec.~\ref{sec:3spin}.
Phase interactions that do not involve the system spins can be safely ignored in the time evolution operator as they commute with all other terms in the Hamiltonian and the $N_B$ environment spins will be traced out in the end.

The phases arising from pairwise interactions between any two spins can be encoded in the adjacency matrix $\underline{ \underline{\Gamma}}(t)$ of a weighted graph whose vertices represent the individual spins. They are connected by an edge whenever they have interacted in the past of $t$, to which we assign the acquired pairwise phase as the weight, $\Gamma_{kn}(t) = \int_0^t \diff t' \tilde{g}_{kn} (t') = \Gamma_{nk}(t)$ for $k\neq n$ and $\Gamma_{kk} (t) = 0$. Thus the adjacency matrix elements describe the interaction history between spins $k$ and $n$ \cite{Hartmann_2007,Hein2004}. A basis vector $|\underline{ab}\ra $ accumulates the two-body phase
\begin{equation}
    \begin{split}
    \oU_t^{(2)} |\underline{ab} \ra &= e^{- \frac{i}{2} \underline{ab} \cdot \underline{\underline{\Gamma}}(t) \cdot \underline{ab} } |\underline{ab} \ra \\
    &= e^{-\frac{i}{2} \left[ \underline{a} \cdot \underline{\underline{\Gamma_{AA}}}(t) \cdot \underline{a} + 2 \underline{a} \cdot \underline{\underline{\Gamma_{AB}}}(t) \cdot \underline{b} + \underline{b} \cdot \underline{\underline{\Gamma_{BB}}}(t) \cdot \underline{b} \right]} |\underline{ab} \ra .
    \end{split}
\end{equation}
Here, terms of the form $\underline{\underline{\Gamma}}\cdot \underline{v}$ and $\underline{w}\cdot\underline{v}$ denote the usual real-valued matrix-vector and  inner product, respectively. The factor $1/2$ ensures that each pair is counted once. 
For the reduced system state evolution, we only need to consider the submatrix $\underline{\underline{\Gamma_{AB}}} (t)$ of the $N_A\times N_B$ phases coupling system and environment spins as well as the intra-system couplings, generally denoted by the $N_A \times N_A$ matrix $\underline{\underline{\Gamma_{AA}}} (t)$. Here we consider one fixed coupling strength $g_{AA}$ between the $N_A=2$ system spins, i.e., 
\begin{equation}
    \underline{\underline{\Gamma_{AA}}} (t) = \begin{pmatrix} 0 & g_{AA} t \\ g_{AA} t & 0 \end{pmatrix},
\end{equation}
but our model could be straightforwardly extended to include more spins as part of the $A$-system. 

Incorporating arbitrary three-spin interactions will complicate the graph model substantially as it demands we introduce hyperedges between three vertices and keep track of a much larger adjacency matrix that records the interaction history between spin triples. Given that physical three-body interactions are typically short-ranged and weak, we shall restrict our view to spin triples in which two out of three pairs are nearest neighbours, to which we assign at a fixed coupling strength $h$. (For simplicity, we assign $2h$ if all three spins are mutual nearest neighbours.) 
To model this, let us introduce a second time-dependent adjacency matrix $\underline{\underline{\Lambda}} (t)$ whose binary matrix elements $\Lambda_{kn} (t) \in \{0,1\}$ indicate whether the spins $k$ and $n$ are neighbours at the given time step $t$.

For a system of interest consisting of $N_A=2$ neighbouring spins, we can distinguish two relevant types of mutual neighbour triples: (i) both system spins and one environment spin, and (ii) one system spin and two environment spins. The associated three-body phases can be expressed in terms of the submatrices $\underline{\underline{\Lambda_{AB}}} (t)$ and $\underline{\underline{\Lambda_{BB}}} (t)$ indicating system-environment and environment-environment neighbours at each time, respectively,
\begin{equation}
    \begin{split}
    &\oU_t^{(3)} |\underline{ab} \ra = \\
    & e^{-ih \int_0^t \diff t' \left\{ a_1 a_2 \, (1,1) \cdot \underline{\underline{\Lambda_{AB}}}(t') \cdot\underline{b}  \,+ \,  \underline{a} \cdot \underline{\underline{\Lambda_{AB}}}(t') \cdot \left[ \underline{b} \circ \underline{\underline{\Lambda_{BB}}}(t') \cdot \underline{b} \right] \right\} } |\underline{ab} \ra .
    \end{split}
\end{equation}
Here, the $\circ$-operator stands for the Hadamard product, i.e., elementwise multiplication of two arrays or matrices of equal shape. The first term in the exponent represents type (i), which only contributes for $\underline{a}=(a_1,a_2)=(1,1)$.

In our model, we assume that we have full control over the $N_A$ system spins, which we can initialize in some pure fiducial state $\rho_A = |\psi_A\ra \la \psi_A |$, while the environment is in a given (uncontrolled, e.g., thermal) state $\rho_B$. Clearly, when we trace over the $N_B$ environment spins, the reduced system state evolution will depend only on the diagonal elements $p(\underline{b}) = \la \underline{b} | \rho_B | \underline{b} \ra$, and we need not care about coherences. 
We can thus assume the global initial state
\begin{equation}
    \rho_{AB}(0) = \sum_{\underline{a},\underline{a}';\underline{b}} \la \underline{a}|\rho_A|\underline{a}'\ra p(\underline{b}) |\underline{a}\ra \la \underline{a}'| \otimes |\underline{b}\ra\la \underline{b}|.
\end{equation}
This state evolves under the general Hamiltonian \eqref{eq:Halt} for a time $t$, after which we obtain the reduced two-spin system state as

\begin{equation}
    \begin{split}
    \rho_A(t) &= \tr_B \left[ \oU_t \rho_{AB} (0) \oU_t^\da \right] \\
    &= \sum_{\underline{a},\underline{a}'} \la \underline{a}|\rho_A|\underline{a}'\ra  C_t (\underline{a},\underline{a}')e^{ig_{AA}t (a_1'a_2'-a_1a_2)} |\underline{a}\ra \la \underline{a}'|,
    \end{split}
\end{equation}
with the decoherence factor
\begin{equation}
    \begin{split}
       C_t (\underline{a},\underline{a}') = & \sum_{\underline{b}} p(\underline{b}) e^{i \left(\underline{a}'-\underline{a}\right) \cdot \underline{\underline{\Gamma_{AB}}}(t)\cdot \underline{b}} \\
        & \times e^{ih \int_0^t \diff t' (a_1' a_2' - a_1 a_2) \, (1,1) \cdot \underline{\underline{\Lambda_{AB}}}(t') \cdot\underline{b} } \\ 
        & \times e^{ih \int_0^t \diff t' (\underline{a}' - \underline{a}) \cdot \underline{\underline{\Lambda_{AB}}}(t') \cdot \left[ \underline{b} \circ \underline{\underline{\Lambda_{BB}}}(t') \cdot \underline{b} \right] } . \label{eq:decoherenceFactor}
        \end{split}
\end{equation}
In the next sections, the decoherence factor matrix of reduced system states is given out directly for specified initial state of system, environment and Hamiltonian of global system.

\section{2D Lattice structure with short-range interaction}\label{sec:lattice}

In our first case study, we are concerned with the difference between Markovian and non-Markovian dephasing that deteriorates the fidelity and entangling power of a phase gate between the system spins. We will also discuss how random reset operations can preserve some degree of entanglement at long times, and how three-body interactions can mitigate the dephasing effect.

We consider a two-dimensional lattice model based on a Hamiltonian of the form \eqref{eq:H} with $\alpha=0$, which resembles a generic Rydberg array of moderate size. Two separate and strongly interacting atoms constitute the system, which exchanges phase information through nearest-neighbour Rydberg blockade interactions of strength $g_{AB} > 0$ with an environment of $N_B=100$ atoms distributed over the lattice. We assume a worst-case scenario of an infinite-temperature environment with uniform excitation probability $p(\underline{b}) = 2^{-N_B}$, equivalent to the initial state $\rho_B = |+\ra\la +|^{\otimes N_B}$ used in Refs.~\cite{Calsamiglia2005, Hartmann2005}. 

Starting from a uniformly filled lattice with one $B$-atom per site, environmental fluctuations can be emulated by letting the atoms hop to random neighbouring sites at a given rate $\eta_B >0$. For simplicity, we assume that one site can be occupied by multiple atoms and that the lattice topology is that of a torus with periodic continuation at its boundaries. All atoms on the same and on the eight surrounding sites are counted as nearest neighbours with a fixed coupling rate $g_{AB} \Delta t$. 
Two opposing regimes will be compared in the following: (i) a strongly non-Markovian regime where the two system atoms are placed on top of the $B$-atoms at neighbouring sites in the lattice center and remain there, and (ii) an almost Markovian regime in which the location and thus the immediate surrounding environment of the two system spins is switched randomly at each time step of the simulation. 

In the absence of three-body interactions, the decoherence factor \eqref{eq:decoherenceFactor} for the infinite-temperature lattice environment simplifies greatly to
\begin{equation}
    \begin{split}
    C_t (\underline{a},\underline{a}') = &  \prod_{n=1}^{N_B} e^{\frac{i}{2} \left(\underline{a}'-\underline{a}\right) \cdot \underline{\underline{\Gamma_{AB}}}(t)\cdot \underline{b}^{(n)}} \\
    &\times \cos \left[\frac{1}{2} \left(\underline{a}'-\underline{a}\right) \cdot \underline{\underline{\Gamma_{AB}}}(t)\cdot \underline{b}^{(n)} \right] , \label{eq:decoherenceFactor_2body_lattice}
    \end{split}
\end{equation}
where the matrix of pairwise phases is now determined by the neighbour adjacency matrix between system and environment spins,  $\underline{\underline{\Gamma_{AB}}} (t) = g_{AB} \int_0^t \diff t' \, \underline{\underline{\Lambda_{AB}}}(t')$. The $\underline{b}^{(n)}$ denote the $N_B$ basis arrays with only the $n$-th atom in the $|1_n\ra$ state and all others in $|0\ra$. The decoherence factor \eqref{eq:decoherenceFactor_2body_lattice} can be computed efficiently for large environments, $N_B \gg 1$, and will be the starting point for our following assessment of an entangling phase gate under environmental dephasing. Our simulations are performed using discrete time steps $\Delta t$ and random hopping events of the environment spins at every $1/\eta_B$-th step.

\subsection{Entangling power of a two-qubit phase gate under dephasing}\label{sub:two_qubit_gate}

For our case study, we consider two system spins initialized in the product state $|\psi_A\ra = |+\ra|+\ra$, which remain in close vicinity at a fixed interaction rate $g_{AA} \Delta t$. In isolation, the Rydberg interaction will build up entanglement periodically, resulting in a maximally entangled Bell state at odd multiples $g_{AA}t = (2n+1)\pi$. The interaction thus implements a perfect controlled phase (c-phase) gate, which can facilitate universal quantum computing with qubit arrays \cite{Saffman2010}. However, the presence of other surrounding spins causes dephasing and thus quickly diminishes the gate fidelity. A good indicator for this effect is the amount of system entanglement over time, measured in terms of the negativity \cite{Vidal2002,Plenio2005},
\begin{equation}
    \mathcal{N} (t) = \frac{\lVert \rho^{T_{A}} \rVert - 1}{2}, 
    \label{eq:negativity}
\end{equation}
which ranges from zero to its maximum value $0.5$.

\begin{figure}[t]
    \centering
    \includegraphics[width=8.5cm]{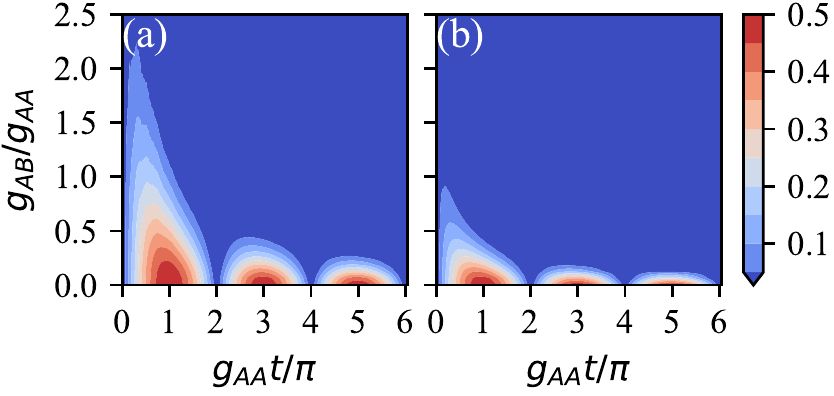}
    \caption{Entanglement negativity between the system spins versus time $g_{AA}t/\pi$ and system-environment coupling $g_{AB}/g_{AA}$. The spins get entangled through their phase interaction $g_{AA}\Delta t = 0.05$, while subject to (a) quasi-Markovian and (b) non-Markovian dephasing from an infinite-temperature spin environment. The hopping rate of environmental spins is $\eta_B =1/5$, and in (a) the system spins randomly jump between lattice sites at every time step $\Delta t$. We evaluate the negativity for ensemble-averaged states over 200000 trajectories.\label{fig:neg_densityplot}}
\end{figure}

In Fig.~\ref{fig:neg_densityplot}, we compare the impact of (a) Markovian and (b) non-Markovian dephasing on the achieved c-phase gate negativity as a function of time $t$ and relative environment coupling strength $g_{AB}/g_{AA}$. We averaged the system state $\rho$ over 200000 simulated trajectories on an exemplary $10\times 10$ lattice filled uniformly by $N_B=100$ spins that then hop to random neighbouring sites every $5$ time steps. Panels (a) and (b) correspond, respectively, to the almost Markovian regime in which the two system spins randomly relocate in every time step and to the non-Markovian limit in which the system spins stay at a fixed position inside the lattice (both spins on the third site from the left and second site from the top). 

Clearly, the non-Markovian regime in (b) leads to a more severe entanglement decay: negativities greater than $0.1$ are confined to weak coupling strengths $g_{AB} \lesssim 0.5 g_{AA}$ and the periodic recurrence of entanglement is barely visible. Notice also that the time of maximum entanglement decreases with growing environment coupling, which is a consequence of an effective Lamb shift of the system energies. We conclude that, at a given two-spin coupling strength, the phase information and entanglement in the system are more stable in a dynamic Markovian environment than in a static non-Markovian one.

One may ask whether the observed behaviour and discrepancy could be modeled in terms of strictly Markovian dephasing channels with rescaled effective coupling rates. To this end, we consider a discrete-time Markov chain describing dephasing of the two system spins subject to the same local environment in the interaction picture. At the $j$-th time step $\Delta t$, the system state updates to $\rho_{A,j} = \rho_A (t=j\Delta t)$ according to
\begin{equation}
    \rho_{A,j} = \tr_B\left[ \oV_{AB} \oU_A \left( \rho_{A,j-1} \otimes \left(\frac{\openone}{2}\right)^{\otimes n_B}\right) \oU_A^\da \oV_{AB}^\da \right], \label{rho_disc}
\end{equation}
where the system evolution is described by the unitary $\oU_A = e^{i g_{AA} \Delta t |1_1 1_2 \ra \la 1_1 1_2|}$ and the system-environment coupling by
\begin{equation}
    \oV_{AB} = \prod_{n=1}^{n_B} e^{-i g_{AB}\Delta t (|1_1\ra \la 1_1| + |1_2\ra \la 1_2|) \otimes |1_n \ra \la 1_n|}.
\end{equation}
The update rule \eqref{rho_disc} describes the discrete-time dynamics of the system in the framework of collision models \cite{Francesco2022}, assuming both system spins interact with an independent bunch of $n_B$ maximally mixed environment spins in each step. 
In the computational basis $|\underline{a}\ra$, we can rewrite the Markov update rule in terms of a Hermitean coefficient matrix 
\begin{equation}
    \la \underline{a} | \rho_{A,j} | \underline{a}'\ra  = C_M (\underline{a},\underline{a}') \la \underline{a} | \rho_{A,j-1} | \underline{a}'\ra . \label{Coef_Mat}
\end{equation}
The diagonal elements of this matrix are all $C_M (\underline{a},\underline{a})=1$, while the off-diagonal elements are given by
\begin{equation}
    \begin{split}
        C_M (00,01) &= C_M (00,10) \\
        &= e^{i \frac{1}{2} n_B g_{AB}\Delta t} \cos^{n_B}{\left(\frac{g_{AB}\Delta t}{2} \right)}, \\
        C_M (00,11) &= e^{i (g_{AA} + n_B g_{AB})\Delta t} \cos^{n_B}{\left(g_{AB}\Delta t\right)}, \\
        C_M (01,10) &= 1, \\
        C_M (01,11) &= C_M (10,11) = e^{i g_{AA}\Delta t} C_M (00,01),
    \end{split}
\end{equation}
and the respective Hermitean conjugates.
A straightforward calculation starting from the initial product state then yields the entanglement negativity in each time step.

\begin{figure}[t]
    \centering
    \includegraphics[width=\linewidth]{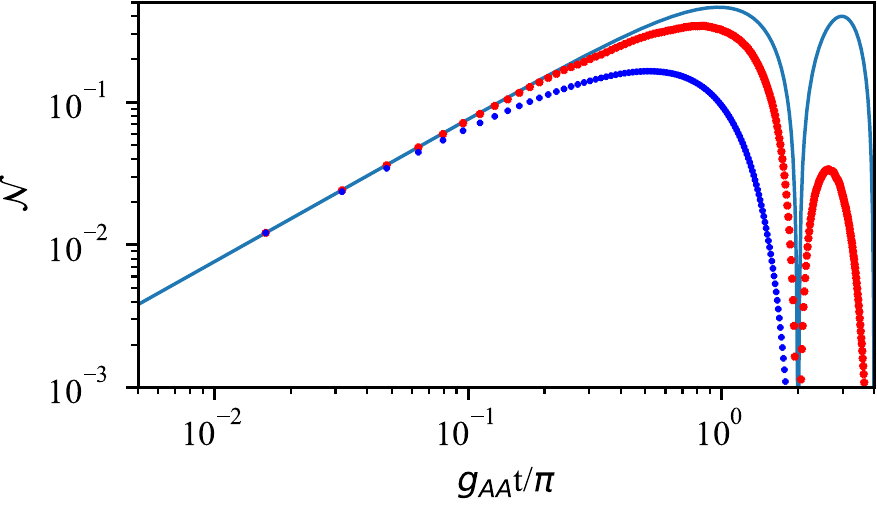}
    \caption{Entanglement negativity as a function of time at $g_{AA} \Delta t=0.05$ and $g_{AB}/g_{AA}=0.5$. We compare the quasi-Markovian (red stars) and the non-Markovian case (blue dots) from Fig.~\ref{fig:neg_densityplot} against the strictly Markovian collision model based on \eqref{Coef_Mat} with $n_B = 8$ (solid line). We take the ensemble average over $10^5$ trajectories here.}\label{fig:neg_ME_fit}
\end{figure}

In Fig.~\ref{fig:neg_ME_fit}, we compare the simulation results for the negativity decay from Fig.~\ref{fig:neg_densityplot} at fixed $g_{AB}=0.5g_{AA}$ and $g_{AA}\Delta t = 0.05$ to the Markov collision model \eqref{Coef_Mat} with $n_B=8$ (solid line). The latter amounts to the average expected number of neighbouring environment spins in a fully filled lattice. 
The results from the quasi-Markovian regime (red dots) agree well with the Markov model for about the first 15 random hopping steps, after which the finite lattice size leads to growing deviations. The Markov rule \eqref{Coef_Mat} keeps subjecting the system to an \emph{independent} set of $n_B$ environment spins at every step, whereas in the actual simulation, the system spins start seeing the same neighbour particles again after a while, breaking the Markov assumption. 
Consistently, the negativity associated to the static non-Markovian simulation (blue dots) decays more rapidly and thus deviates much earlier from the simple collision model.

\subsection{Reset-based entanglement preservation} 

In the previous subsection, we have studied the entangling power of a c-phase gate between two spins subject to dephasing from surrounding spins. We observed that a dynamically changing, quasi-Markovian local environment is less detrimental than a static non-Markovian one, but in any case the entanglement (and the gate fidelity) decay exponentially on the short time scales $g_{AB}t \lesssim 1$ determined by the system-environment coupling. 
Here we discuss a simple repeated intervention method to generate and uphold a non-vanishing average amount of entanglement for on-demand use at arbitrary long times. 

Consider an operation that resets the two system spins to their initial or any other pure product state $|\psi_1\psi_2\ra$, which is a CPTP map with Kraus representation
\begin{equation}
    \begin{split}
    \Phi_r (\rho_A) &= |\psi_1\psi_2\ra\la \psi_1 \psi_2| \\
    &= \sum_{a_1,a_2=0}^1 |\psi_1\psi_2 \ra \la a_1 a_2| \rho_A |a_1 a_2 \ra \la \psi_1\psi_2 |.
    \end{split}
\end{equation}
By repeating this local operation at regular or random times with an average rate $\kappa \Delta t$ that is comparable to the environment coupling strength, one can partially counteract the dephasing and make the system equilibrate to an ensemble-averaged steady state with a finite amount of entanglement at times $\kappa t > 1$. To this end, the reset states should lie on the equator of each spin's Bloch sphere.

To demonstrate the effect, we performed simulations implementing a Poisson process of reset interventions at various rates $\kappa \Delta t$ in our lattice dephasing model: For each trajectory, the waiting times until subsequent reset events are drawn from an exponential distribution and the system is evolved according to our decoherence model \eqref{eq:decoherenceFactor_2body_lattice} in between the events. As expected, we found that the most entanglement is retained when resetting to the initial state $|\psi_1 \psi_2\ra = |++\ra$. 

\begin{figure}[t]
    \centering
    \includegraphics[width=\linewidth]{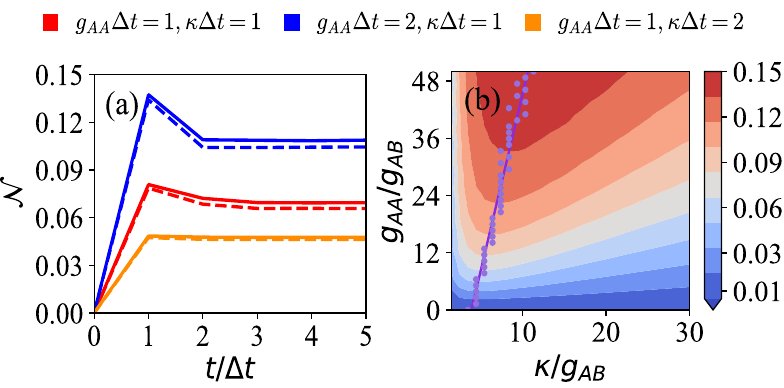}
    \caption{(a) Entanglement negativity versus time in the presence of dephasing and random simultaneous reset operations on the two system spins. The solid and dashed lines represent the quasi-Markovian and the non-Markovian case, respectively; the colors represent different combinations of intra-system coupling and reset rate. (b) Steady-state negativity versus reset rate $\kappa$ and intra-system coupling $g_{AA}$, as evaluated at $t=15\Delta t$. The dots show the maximum negativity with respect to $\kappa$ at fixed $g_{AA}$, the line denotes $\kappa=7g_{AA}$. We fix $g_{AB}\Delta t = 0.1$ and average over $2\times 10^5$ trajectories.}\label{fig:neg_with_reset}
\end{figure}

Figure \ref{fig:neg_with_reset} shows the results averaged over 200000 simulated trajectories per parameter set. Panel (a) shows the quick equilibration to steady-state entanglement negativity as a function of time for a few exemplary intra-system coupling rates $g_{AA}/g_{AB}$ and reset rates $\kappa/g_{AB}$. The quasi-Markovian environment (solid) and the static non-Markovian environment (dashed) yield approximately the same results here. In panel (b), the steady-state negativity is evaluated as a function of the two rates, showing that one achieves the maximum entanglement when resetting at a rate $\kappa \sim 7 g_{AA}$. Generally, at weak system-environment couplings, $g_{AA}\gg g_{AB}$, a significant fraction of entanglement can be preserved on average.
A similar, but less pronounced entanglement-preserving effect was noticed previously \cite{Hartmann2006} for random single-spin reset interventions, starting from a maximally entangled state.

The reset protocol we consider here merely requires synchronised single-qubit control at a rate comparable to the c-phase gate time scale, which should be feasible on current spin array platforms for quantum information processing \cite{Saffman2010, Bloch2012, Monroe2021}.

\subsection{Three-body interaction curbing dephasing} \label{sec:3spin}

We now discuss the impact of three-spin interactions on the entanglement decay under dephasing. In a physical scenario, such interactions could be repulsive or attractive, but we would expect them to be significantly weaker than the two-body terms. We will proceed to show that weak three-body couplings of opposite sign compared to the two-body terms can already alleviate the dephasing effect and drastically improve the achievable entanglement and fidelity of a c-phase gate between two system spins.

\begin{figure}[htbp]
    \centering
    \includegraphics[width=\columnwidth]{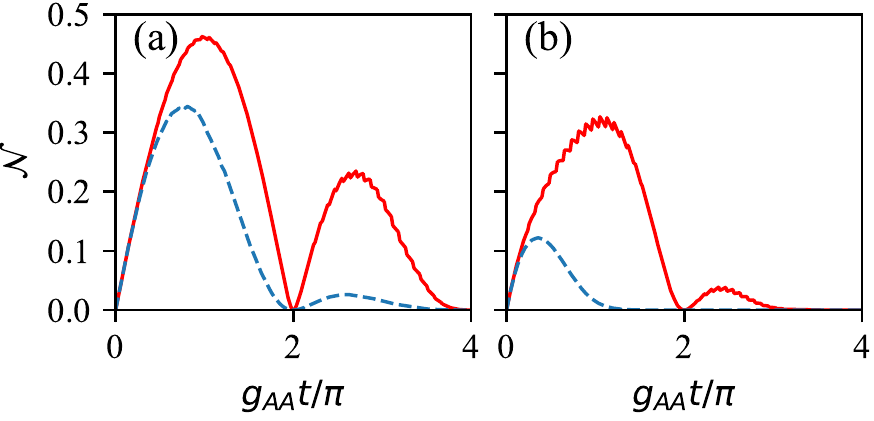}
    \caption{Influence of three-body interactions on the entanglement negativity decay. We plot the negativity as a function of time for the same settings as in Fig.~\ref{fig:neg_ME_fit} (dashed line), with an additional three-body coupling of strength $h = -0.015g_{AB}$ (solid). Once again, we compare (a) the quasi-Markovian to (b) the non-Markovian case, averaged over 10$^5$ trajectories. 
    }\label{fig:neg_3body_effect}
\end{figure}

In our exemplary lattice model, we consider three-body interactions of fixed rate $h<0$ among nearest neighbours only. Given the worst case of maximally mixed environment spins, the decoherence factor \eqref{eq:decoherenceFactor} reduces once again to an efficiently computable product of $N_B$ terms,
\begin{equation}
 \begin{split}
    & C_t (\underline{a},\underline{a'}) = \\
    & \prod_{n=1}^{N_B} e^{\frac{i}{2} (\underline{a}-\underline{a'}) \cdot \left[ \underline{\underline{\Gamma_{AB}}}(t) \cdot \underline{b}^{(n)} +  h \underline{\underline{\Lambda_{AB}}}(t) \cdot \left( \underline{b}^{(n)} \circ \underline{\underline{\Lambda_{BB}}}(t) \cdot \underline{b}^{(n)} \right) \right]} \\
    & \cos \Bigg( \frac{1}{2} (\underline{a}-\underline{a'}) \cdot \Bigg[ \underline{\underline{\Gamma_{AB}}}(t) \cdot \underline{b}^{(n)} +  h \underline{\underline{\Lambda_{AB}}}(t) \cdot \\
    &\Bigg( \underline{b}^{(n)} \circ \underline{\underline{\Lambda_{BB}}}(t) \cdot \underline{b}^{(n)} \Bigg) \Bigg] \Bigg). \\
    \end{split}
\end{equation}
Figure \ref{fig:neg_3body_effect} compares the c-phase entanglement negativity with and without the three-body coupling for a simulated average over 100000 trajectories in our lattice model. Panels (a) and (b) correspond to the quasi-Markovian and the static non-Markovian regime, respectively. We chose the same fixed two-body environment coupling $g_{AB} = 0.5 g_{AA}$ as in the previous Fig.~\ref{fig:neg_ME_fit}, and the blue curves here match the simulated results there. An additional weak three-body coupling of strength $h=-0.015g_{AB}$ results in much higher reachable entanglement values, as depicted by the red curves. The improvement is more pronounced in the non-Markovian case (b) and comes with a notable shift to longer gate times at which maximum entanglement is achieved. Hence, engineering such counteracting three-spin interactions in practice could stabilise local gate operations on qubit array platforms for quantum computing.

Systems of trapped Rydberg ions (or atoms) provide a viable testbed for demonstrating and assessing the effect experimentally, given the high degree of control over both electronic and vibrational degrees of freedom via external laser fields. This not only facilitates tunable interactions between excited Rydberg states by virtue of the Rydberg blockade, but the phonon-mediated dipole-dipole coupling between neighbouring Rydberg states can also lead to an effective three-body anti-blockade interaction \cite{Gambetta20201, Gambetta2020}. By tuning the ratio of two-dimensional trapping frequencies in the $x y$-plane, $\alpha = \omega_y/\omega_x$, one flexibly varies the relative strength and sign of the effective two- and three-body interactions between neighbouring Rydberg ions.

\section{Dephasing in a correlated finite bath at finite temperatures} \label{sec:molecule}

In the previous section, we have studied the effect of non-Markovian spin dephasing on entangling gates embedded in a bath resembling a lattice of interacting Rydberg atoms. 
Here, we assess the impact of bath correlations at finite temperatures on the dephasing effect. The required evaluation of the energy spectrum and the partition function of the interacting bath spins increases computation cost drastically and forces us to resort to a smaller bath size, $N_B < 30$. As a physical platform for our case study, we choose NMR spin processing on a single organic molecule. Such a molecule can act as an elementary unit for scalable quantum information processing, given the individually accessible and controllable nuclear spins in the molecule \cite{Anderegg2019,Liu2019,Zhang2020}. However, coherence times for quantum operations are limited by the dephasing caused by dipole-dipole interactions with the other surrounding constituent spins \cite{Zhang2022}. 

The decoherence of a single central spin in a Triphenylphosphine (PPh3)-atom molecule and the concomitant spread of multi-spin correlations has recently been analysed theoretically and verified by measurements \cite{Niknam2021}. The experiment was performed in the high-temperature regime where the environment spins are well described by uncorrelated maximally mixed states. 

\begin{figure}[t]
    \centering
    \includegraphics[width=7cm]{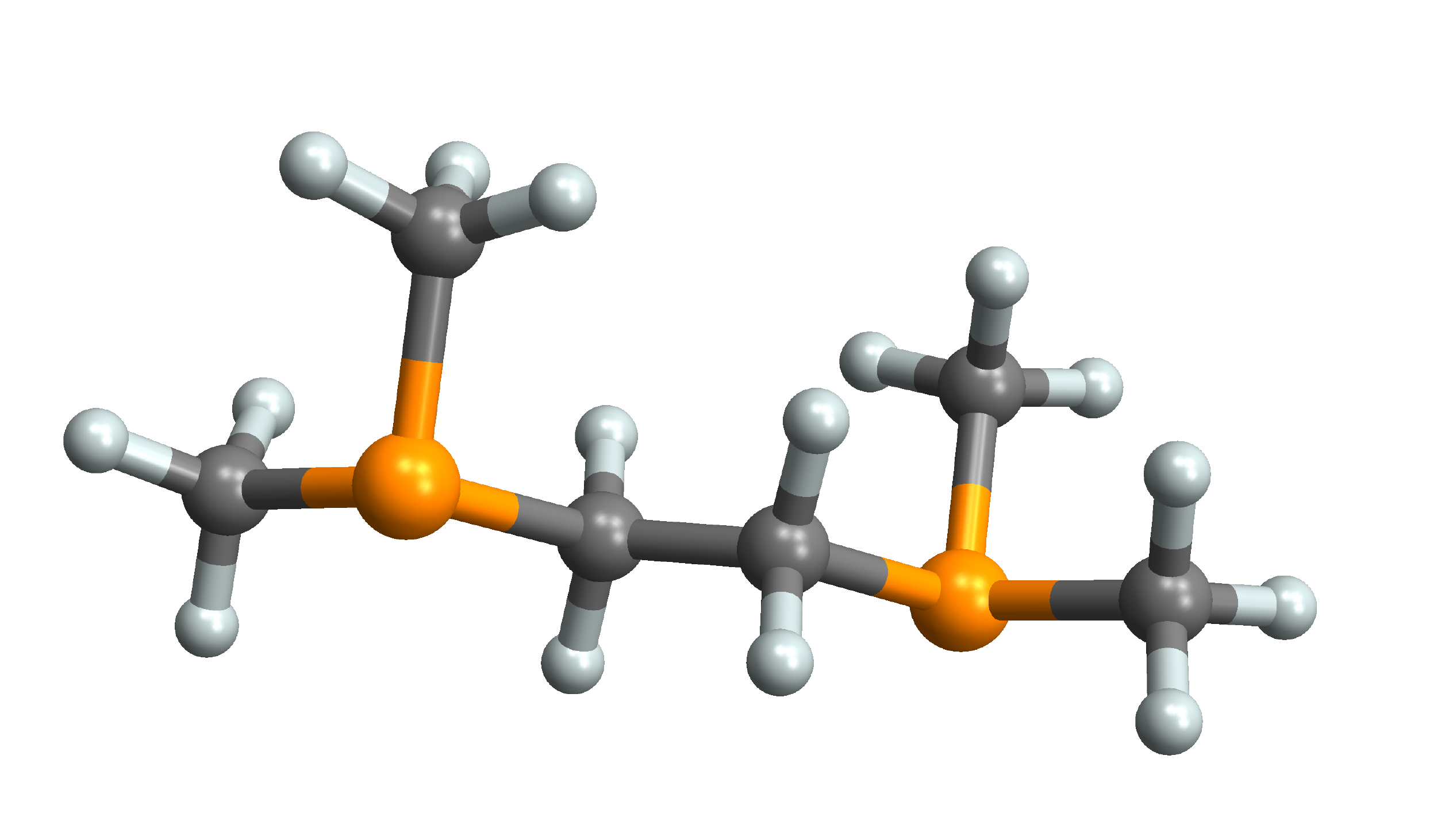}
    \caption{Structure of the 1,2-Bis(dimethylphosphino)ethane (dpme) molecule C$_{6}$H$_{16}$P$_2$. The nuclear spins of the two phosphorus atoms (yellow) act as the quantum system subject to non-Markovian dephasing from the 16 surrounding hydrogen spins (light grey) via magnetic dipole interactions. We average over random orientations of the molecule with respect to the external magnetic field axis.}\label{fig:molecule}
\end{figure}

In our theoretical case study, we will demonstrate that bath correlations will amplify the decoherence effect at low temperatures when bath correlations are non-negligible. We shall employ a smaller, computationally easier to handle molecule for our considerations: the 1,2-bis(dimethylphosphino)ethane (dpme) structure C$_{6}$H$_{16}$P$_2$, as depicted in Fig.~\ref{fig:molecule}. The nuclear spins of the two central phosphorus atoms will act as our system of interest, while the $N_B=16$ hydrogen nuclear spins constitute the thermal environment. 

In order to employ our framework and to observe a pronounced influence of correlations, we shall make two crucial assumptions: First, we consider nuclear spins in thermal equilibrium at very low temperatures down to the nK scale, as opposed to present-day NMR experiments typically operated at room temperature. Second, we simplify the dipole-dipole interactions between the spins by an Ising-type coupling, which results in pure dephasing, but is strictly valid only at high magnetic field strengths \cite{Duer2001}.

The Hamiltonian for the spin system is thus of the form \eqref{eq:Halt} with $\alpha=1$, where the quantisation axis $\bm e_z$ for the spin states $|0\ra,|1\ra$ is set by an external homogeneous magnetic field of strength $B_z$. The corresponding bare resonance frequencies for the hydrogen spins are all equal to $\omega_{\rm H} = g_{\rm H} \mu_N B_z/\hbar$ with $g_{\rm H} = 5.585$ the hydrogen g-factor and $\mu_N$ the nuclear magneton. The magnetic dipole-dipole interaction between any two spins yields the two-spin coupling frequencies 
\begin{equation}
    \tilde{g}_{kn} = g(\bm r_k,\bm r_n) = \frac{\mu_0 \mu_N^2 g_k g_n}{4\pi \hbar |\bm r_k - \bm r_n|^3} \left[1 - 3 \frac{(z_k-z_n)^2}{|\bm r_k-\bm r_n|^2} \right], \label{eq:dipoleCoupling_g}
\end{equation}
where the values for the g-factors are $g_{\rm H}$ and $g_{\rm P} = 2.261$ for the environment spins and the system spins, respectively. We consistently set $\tilde{g}_{kk}=0$; three-body couplings are omitted. Notice that orientation of the molecule enters here as the coupling between any two spins depends on the angle between their distance vector and the magnetic field axis. 

In NMR experiments, one typically employs a powder of the molecular substance at cold temperatures. We shall therefore consider an ensemble of molecules with randomly distributed static orientations, and the hydrogen spins in each molecule shall be in a thermal equilibrium state at given inverse temperature $\beta = \hbar/k_B T$. The two phosphorus spins are assumed to be prepared in the maximally entangled state $|\psi_A\ra=\frac{1}{\sqrt{2}}(|00\ra + |11\ra)$, which is now subjected to dephasing by the hydrogen spin bath.

The thermal state of the hydrogen spins and its partition function read as
\begin{equation}
    \begin{split}
    \rho_B &= \sum_{\underline{b}} p(\underline{b}) |\underline{b}\ra\la \underline{b}| \\
    & =  \frac{1}{Z_B(\beta)}\sum_{\underline{b}}e^{-\beta \left(\underline{\tilde{\omega}_B}\cdot \underline{b} + \frac{1}{2} \underline{b}\cdot \underline{\underline{\tilde{g}_B}} \cdot \underline{b}\right)}|\underline{b}\ra \la \underline{b}|, \label{eq:GibbsState} \\
    Z_B(\beta) &= \tr{(e^{-\beta \oH_B})} = \sum_{\underline{b}} e^{-\beta\left(\underline{\tilde{\omega}_B}\cdot \underline{b} + \frac{1}{2} \underline{b}\cdot  \underline{\underline{\tilde{g}_B}}\cdot \underline{b} \right)}, 
    \end{split}
\end{equation}
where $\underline{\underline{\tilde{g}_B}}$ is the $16\times 16$ matrix of homonuclear H-H coupling frequencies \eqref{eq:dipoleCoupling_g} with zeros on the diagonal. The renormalized hydrogen eigenfrequencies in $\underline{\tilde{\omega}_B}$ are given by \eqref{eq:rescaled_w_g} as
\begin{equation}
  \tilde{\omega}_k = \omega_{\rm H} - \frac{1}{2} \sum_{n_A = 1}^2 g(\bm r_k,\bm r_{n_A}) - \frac{1}{2} \sum_{\substack{n = 1 \\ n\neq k}}^{16} g(\bm r_k,\bm r_n). \label{eq:NMR_eigenfreq_renorm}
\end{equation}
They include the heteronuclear coupling contributions from the two P-spins and the homonuclear ones from the other H-spins. The orientation-independent prefactors in \eqref{eq:dipoleCoupling_g} give the magnitude of the coupling frequencies, which range from $2\pi \times 27.5\,$Hz to $2\pi \times 3.52\,$kHz here, while the bare H-spin resonance $\omega_{\rm H}$ amounts to about $2\pi \times 42.5 \,$kHz/mT. Hence bath correlations are only significant at weak magnetic fields, $B_z \lesssim 0.1\,$mT. Whereas, at high magnetic fields, the local spin energies are dominant and we can approximate the thermal state by a product of local Gibbs states,
\begin{equation} \label{eq:thermalProbUncorr}
    p(\underline{b}) \approx \prod_{n=1}^{N_B} \frac{e^{-\beta \tilde{\omega}_n b_n}}{1 + e^{-\beta \tilde{\omega}_n}} \approx \frac{e^{-\beta \omega_{\rm H} \underline{b}\cdot\underline{b}}}{\left( 1 + e^{-\beta \omega_{\rm H}}\right)^{N_B}},
\end{equation}
regardless of the molecular orientation. 

In general, the thermal populations $p(\underline{b})$ and thus the decoherence factor \eqref{eq:decoherenceFactor} for the P-spins depend also on the molecule orientation through the P-H coupling terms \eqref{eq:dipoleCoupling_g}. To implement an ensemble of uniformly random static orientations, each in thermal equilibrium, we average the decoherence factor uniformly over a sphere defined by the two relevant Euler angles: the polar angle $\theta \in [0,\pi]$ with respect to the field $z$-axis and the rotation angle $\gamma \in [0,2\pi]$ around the principal axis of molecule. We arrive at
\begin{equation} \label{molecular_dec_factor}
        C_t (\underline{a},\underline{a}') = \iint \frac{\sin\theta}{4\pi} \diff \theta \diff\gamma \sum_{\underline{b}} p(\underline{b}|\theta,\gamma) e^{i t \left(\underline{a}-\underline{a}'\right) \cdot \underline{\underline{\tilde{g}_{AB}}} ( \theta, \gamma ) \cdot \underline{b}} .
\end{equation}
Notice that, as the bath state is correlated here, the sum over all $2^{N_B}$ binary vectors $\underline{b}$ no longer reduces to a product of $N_B$ terms as in our lattice model. The computational cost thus grows exponentially with the bath size. 

\begin{figure}[tbp]
    \centering
    \includegraphics[width=\linewidth]{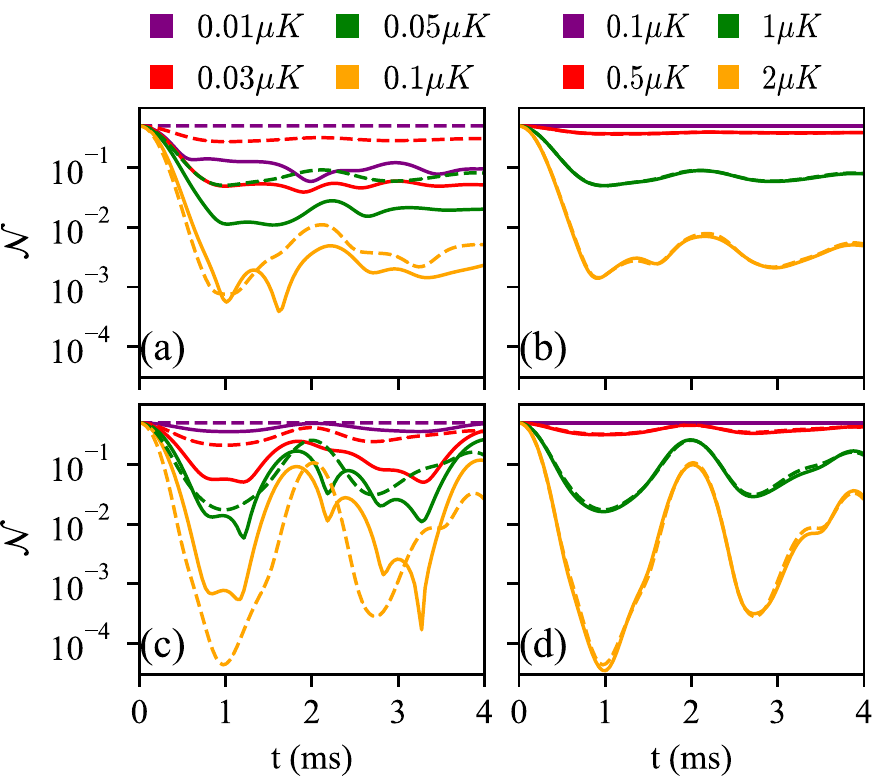}
    \caption{Entanglement negativity decay with time, for two initially Bell-entangled nuclear P-spins under dephasing by the surrounding H-spins of the dpme molecule at a magnetic field of (a,c) $B_z=0.05\,$mT, and (b,d) $1\,$mT. The molecule is (a,b) at a fixed orientation corresponding to $\theta = \pi/2, \gamma = 0$ and (c,d)  averaged over 5000 random orientations with respect to the magnetic field axis. We compare the dephasing in an interacting (solid) and a non-interacting (dashed) thermal environment at various temperatures.}\label{fig:moleculeNegDecay}
\end{figure}

Figure \ref{fig:moleculeNegDecay} shows the bath-induced decay of entanglement negativity between the two P-spins as a function of time, for a fixed orientation (a,b) and averaged over 5000 random orientations of the dpme molecule with respect to the magnetic field axis (c,d). We compare the results for a thermal bath state \eqref{eq:GibbsState} with (solid) and without correlations (dashed), where the latter is achieved by setting $\underline{\underline{\tilde{g}_B}}=0$. Curves of different color correspond to different temperatures, panels (a) and (b) correspond to a field strength of 0.05\,mT and 1\,mT, respectively. As expected, the correlated and the uncorrelated bath yield approximately the same result in the strong-field case (b) and also at high temperatures (blue) where the whole bath spectrum is thermally excited. For the weak field in (a), on the other hand, we observe a striking difference in that the correlated bath induces a much more rapid decay at lower temperatures. This is due to the fact that the dipole couplings within the bath generate thermally accessible low-energy excitation modes acquiring phase information from the system. As the thermally excited modes are few in number, pronounced revivals are observed at about $2$ and $4\,$ms of evolution time.

\begin{figure}[tbp]
    \centering
    \includegraphics[width=\columnwidth]{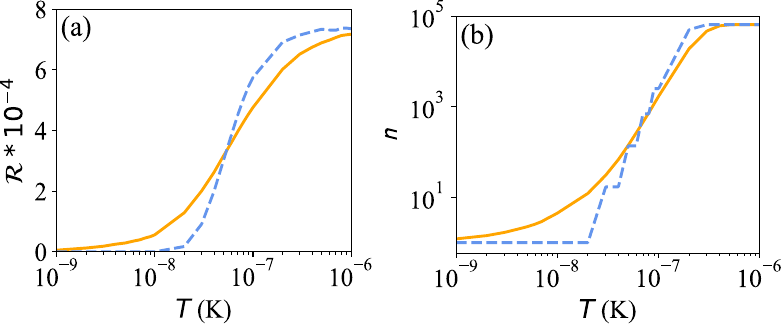}
   \caption{(a) Initial decay $\mathcal{R}$ of entanglement negativity at $\tau = 10\,\mu$s as a function of temperature, comparing the correlated (solid line) to the uncorrelated spin bath (dashed) in the Gibbs state. (b) Number of bath spin states populated by 99\% of the Gibbs distribution. The results were obtained for the weak magnetic field $B_z = 0.05\,$mT and averaged over 5000 random orientations of the molecule.}\label{fig:Initial_Decay_Rate}
\end{figure}

To elucidate the role of bath correlations further, we focus our view on the initial decay of coherence and entanglement. Starting from both system spins prepared in a Bell state with maximum negativity $\mathcal{N}(0)=1/2$, we can expand the decreasing negativity $\mathcal{N}(\tau)$ after a small time step $\tau$. The initial Bell state $|\psi_A\ra$ evolves as
\begin{eqnarray}
    \rho_A(\tau) &=& \frac{1}{2}\Bigg[ |00\rangle \langle 11 | + |11\rangle \langle 00 |  \\
    &&+ C_\tau(00,11)|00\rangle \langle 11 | + C_\tau(11,00)| 11 \rangle \langle 00 |\Bigg], \nonumber
\end{eqnarray}
with the coefficients $C_\tau(\underline{a},\underline{a}')$ obtained from Eq.~(\ref{molecular_dec_factor}). The negativity (\ref{eq:negativity}) of this state assumes the simple form
\begin{equation}
    \mathcal{N}(\tau) = \frac{\sqrt{C_\tau(00,11) C_\tau(11,00)}}{2} = \frac{|C_\tau(00,11)|}{2}.
\end{equation}
Expanding the coefficient $C_\tau$ to second order in $\tau$ results in a quadratic initial decay of negativity,
\begin{eqnarray}
    \mathcal{R} &:=& \mathcal{N} (0) - \mathcal{N}(\tau) \nonumber \\
    &\approx& \frac{1}{2}-\frac{1}{2}\sqrt{1- \tau^2 \var \left[(1,1) \cdot \underline{\underline{\tilde{g}_{AB}}} \cdot \underline{b} \right] }, \nonumber \\
    &\approx& \frac{\tau^2}{4} \var \left[(1,1) \cdot \underline{\underline{\tilde{g}_{AB}}} \cdot \underline{b} \right].
    \label{eq:initialDecayR}
\end{eqnarray}
The variance is taken with respect to the thermal distribution $p(\underline{b}|\theta,\gamma)$ either at a fixed orientation, or including the uniform distribution of orientations. It grows with the number of thermally accessible H-spin states.
For the uncorrelated bath based on the orientation-independent probabilities \eqref{eq:thermalProbUncorr}, the initial decay of negativity factorizes and simplifies to 
\begin{eqnarray}
    \mathcal{R} &\approx&  \frac{\tau^2}{4} \frac{1}{1+e^{\beta \omega_{\rm H}}} \Bigg[ \sum_{n=1}^{N_B}\left< \left( (1,1)\cdot \underline{\underline{\tilde{g}_{AB}}} \cdot \underline{b}^{(n)}\right)^2 \right>_{\theta,\gamma}  \nonumber \\
    & - & \frac{1}{1+e^{\beta \omega_{\rm H}}}\left< \sum_{n=1}^{N_B} (1,1)\cdot \underline{\underline{\tilde{g}_{AB}}} \cdot \underline{b}^{(n)} \right>^2_{\theta,\gamma} \Bigg],
   \label{uncorr_neg.}
\end{eqnarray}
where $\la \cdot \ra_{\theta,\gamma}$ denotes the orientation average.

Figure \ref{fig:Initial_Decay_Rate}(a) shows the initial entanglement decay $\mathcal{R}$ at $\tau=10\,\mu$s as a function of temperature for the correlated (solid) and the uncorrelated H-spin bath (dashed), uniformly averaged over molecular orientations. We find that, at low temperatures, the correlated bath leads to stronger decoherence and entanglement decay, whereas at higher temperatures, the uncorrelated bath takes over. This crossover behaviour can be explained by the number of thermally accessible spin states, because the variance of the P-H coupling in \eqref{eq:initialDecayR} grows the more different H-spin configurations are populated. Panel (b) shows the $99\%$ quantile of thermally occupied spin levels, i.e., the number of spin configurations $\underline{b}$ in ascending order of energy that sum up to $99\%$ of the Gibbs distribution at each given orientation of the uniform average. The crossover region agrees with that of the entanglement decay in (a). The uncorrelated bath exhibits a stepwise increase due to the high degeneracy of its spin states. 

\section{Conclusion} \label{sec:conclusion}

We considered an efficiently solvable, generic decoherence model based on two- and three-body phase interactions in a dynamic network of environmental spins. As a first application, we investigated the entangling power of a controlled phase interaction between two system spins embedded in a periodic spin lattice with nearest-neighbour couplings and random hopping. We showed that the entanglement decays faster under the non-Markovian dephasing caused by the immediate surroundings of the system spins compared to a quasi-Markovian case in which the system spins jump randomly across the lattice. Effective Markovian treatments of dephasing in spin arrays, e.g., in terms of collision models, are therefore inaccurate; they would overestimate the local gate fidelity achievable in spin-based platforms for quantum computing such as Rydberg arrays.

On the other hand, we showed that one can generate steady states with a substantial amount of entanglement by simultaneously resetting both system spins at random times. This improves earlier findings based on single-spin reset operations \cite{Hartmann2006}. The environmental dephasing can also decrease in the presence of three-body interactions. 

Our second case study based on NMR quantum processing with a single molecule spotlights the role of correlations in a thermal spin bath for non-Markovian dephasing. We showed that the correlations could enhance the dephasing at lower temperatures as they give rise to low-energy excitations of the bath, whereas at higher temperatures, the respective uncorrelated bath with its highly degenerate energy subspaces would take over and lead to faster decoherence. Future research could explore local spin measurements as phase probes for correlations and temperature in finite interacting baths \cite{Mitchison2020}.

\acknowledgments
We thank H.~Chau Nguyen and Wolfgang Dür for discussions. 
This work has been supported by the Deutsche Forschungsgemeinschaft (DFG, German Research Foundation, project numbers 447948357 and 440958198), the Sino-German Center for Research Promotion (Project M-0294), the ERC (Consolidator Grant 683107/TempoQ) and the German Ministry of Education and Research (Project QuKuK, BMBF Grant No.~16KIS1618K).


%

\end{document}